\documentclass[10pt,a4paper]{article}

\usepackage{amsmath,amsfonts,amssymb}
\usepackage{mathabx}
\usepackage{stmaryrd}

\usepackage{amsmath}
\usepackage{amsfonts}
\usepackage{epsfig}
\usepackage{graphics} 

\usepackage{tikz}
\usepackage{pgfplots}

\newcommand{\llgk}{\{\hspace*{-0.5ex}\{}
\newcommand{\rrgk}{\}\hspace*{-0.5ex}\}}

\newcommand{\jump}[1]{\llbracket #1 \rrbracket}
\newcommand{\average}[1]{\llgk #1 \rrgk}

\usepackage{authblk}

\begin{document}

\title{Influence of the SIPG penalisation on the numerical properties
of linear systems for elastic wave propagation}


\author[1]{Uwe K\"ocher}



\affil[1]{
Helmut-Schmidt-University, University of the Federal Armed Forces Hamburg,
Department of Mechanical Engineering, Germany,
{\tt koecher@hsu-hamburg.de}
}

\maketitle

\begin{abstract}
Interior penalty discontinuous Galerkin discretisations (IPDG) and especially
the symmetric variant (SIPG) for time-domain wave propagation problems are broadly
accepted and widely used due to their advantageous properties.
Linear systems with block structure arise by applying space-time discretisations
and reducing the global system to time-slab problems.
The design of efficient and robust iterative solvers for linear systems
from interior penalty discretisations for hyperbolic wave equations is still a
challenging task and relies on understanding the properties of the systems.
In this work the numerical properties such as the condition number and
the distribution of eigenvalues of different representations of the linear systems
coming from space-time discretisations for elastic wave propagation
are numerically studied.
These properties for interior penalty discretisations depend on the penalisation
and on the time interval length.
\end{abstract}

\section{Introduction}

The accurate and efficient simulation of time-domain first- and second-order
hyperbolic elastic and acoustic wave propagation phenomena is of importance
in many engineering fields with e.g. electromagnetic, acoustic and seismic
applications as well as for
non-destructive structural health monitoring of light-weighted fibre reinforced
materials; cf. e.g. \cite{koecher_mini_22:Koecher2015,koecher_mini_22:Koecher2014}
and references therein.
Recently there is again an increased interest in the simulation of multi-physics
systems including coupled elastic wave propagation together with fluid flow in
porous media; cf. \cite{koecher_mini_22:Mikelic2012,koecher_mini_22:Biot1940}.
Such models appear for instance in battery engineering and
for biomedical applications.

The ability of the efficient high-order approximation of the space-time wavefield
is of fundamental importance for time-domain numerical simulations of wave phenomena.
Especially discontinuous Galerkin methods (dG) as spatial discretisations exhibit
as favourable over continuous finite element methods in terms of lesser numerical
dispersion at the same discretisation order such as polynomial degree and
degrees of freedom per wavelength; cf. \cite{koecher_mini_22:DeBasabe2008}.
Reed and Hill introduced in 1973 the first discontinuous Galerkin method for
first-order hyperbolic steady-state neutron transport.
Meanwhile there exist many discontinuous Galerkin methods
for the spatial discretisation of elliptic, parabolic and hyperbolic problems;
cf. e.g. \cite{koecher_mini_22:Arnold2002}.

Interior penalty discontinuous Galerkin discretisations (IPDG)
and especially the symmetric interior penalty variant (SIPG) for
time-domain wave propagation problems are broadly accepted and widely used due
to their advantageous properties and
their ability for parallel numerical simulations.
The SIPG discretisation of the second-order wave equations is convergent of
optimal order in the energy- and $L^2$-norm
for any polynomial approximation degree in space;
cf. \cite{koecher_mini_22:Grote2006}.
All interior penalty discontinuous Galerkin families include additional
terms combined of trace operators on the interior and exterior boundaries
between mesh elements.
Interior penalty methods include a stabilisation term which penalises
jumps of the trial and test function traces on the boundaries to ensure
coercivity of the bilinear form corresponding to the Laplacian operator.
The needed weighting depends on spatial mesh parameters such as the cell
diameter, the cell anisotropy and the local polynomial degree as well as on
material parameters but not on the time discretisation parameters; cf.
\cite{koecher_mini_22:Hoppe2008,koecher_mini_22:Grote2006,koecher_mini_22:Arnold2002}.
The estimation of the local minimal choice of the penalisation is of fundamental
importance but it turns out to be difficult in physically relevant problems.
An over-penalisation should be avoided since such results in linear systems with
higher condition numbers than necessary. The over-penalisation may result in
inefficient iterative system solves or breakdowns
if it is not take into account in the solver design.

The common drawback of discontinuous Galerkin discretisations compared to their
continuous Galerkin counterparts is that they need more degrees of freedom to
obtain the same analytic convergence order.
In physically relevant problems the drawback of the higher number of degrees of
freedom is not that present compared to analytic test problems.
To the contrary, physically relevant problems gain advantages,
such as avoiding linear-elasticity locking phenomena and the ability to capture
incompatible but relevant boundary and initial conditions,
from the lesser grid stiffness of discontinuous approaches in space and time;
cf. e.g. \cite{koecher_mini_22:Koecher2015}.

Linear systems with block structure arise by applying space-time discretisations
and reducing the global system to time-slab or time-interval problems;
cf. \cite{koecher_mini_22:Koecher2015}.
The design of efficient and robust iterative solvers for linear systems
from interior penalty discretisations for hyperbolic wave equations is still a
challenging task and relies on understanding the properties of the systems.

This work presents the numerical properties such as the condition number and
the distribution of eigenvalues and their dependency on the penalisation and
on the time interval length of different representations of the linear systems
coming from space-time discretisations for elastic wave propagation.
To the best knowledge of the author such results are not present in the literature.

The following structure of the paper is
the brief introduction of the elastic wave equation and
its space-time discretisation in Sec. \ref{koecher_mini_22:sec:2},
the presentation of different representations of the fully discrete systems
which are further studied in Sec. \ref{koecher_mini_22:sec:3},
the numerical experiments section providing the results in
Sec. \ref{koecher_mini_22:sec:4} and
summarising and concluding remarks in Sec. \ref{koecher_mini_22:sec:5}.

\section{Elastic wave equation and space-time discretisation}
\label{koecher_mini_22:sec:2}

The elastic wave equation in its first-order in time, or displacement-velocity,
representation reads as the following system.
The primal variables are the displacement $\boldsymbol u$ and
the velocity $\boldsymbol v$.
The volume is represented by the domain $\Omega \subset \mathbb{R}^d$,
with dimension $d=2,3$, and
the time domain is given by $I=(0,T)$ for some finite final time $T$.
Find the pair $\{\boldsymbol u,\boldsymbol v\}$ from
\begin{equation}
\label{koecher_mini_22:eq:1:ewave}
\begin{array}{r@{\,}c@{\,}l@{\;}c@{\;}l@{\quad}l}
\rho_s(\boldsymbol x) \partial_t \boldsymbol u(\boldsymbol x,t)
&-& \rho_s(\boldsymbol x) \boldsymbol v(\boldsymbol x,t) &=& \boldsymbol 0 &
\text{in } \; \Omega \times I\,,\\[1.5ex]
\rho_s(\boldsymbol x) \partial_t \boldsymbol v(\boldsymbol x,t)
&-& \nabla \cdot \boldsymbol \sigma(\boldsymbol u(\boldsymbol x,t)) &=&
\boldsymbol f(\boldsymbol x,t) &
\text{in } \; \Omega \times I\,,
\end{array}
\end{equation}
with boundary conditions
$\boldsymbol u(\boldsymbol x,t) = \boldsymbol g(\boldsymbol x,t)$ on
$\Gamma_D \times I$,
$\boldsymbol \sigma(\boldsymbol u(\boldsymbol x,t))\, \boldsymbol n =
\boldsymbol h(\boldsymbol x,t)$ on $\Gamma_N \times I$,
and initial conditions
$\boldsymbol u(\boldsymbol x,0) = \boldsymbol u_0(\boldsymbol x)$ in
$\Omega \times \{0\}$,
$\boldsymbol v(\boldsymbol x,0) = \boldsymbol v_0(\boldsymbol x)$ in
$\Omega \times \{0\}$.
The linearised stress tensor is given by
$\boldsymbol \sigma(\boldsymbol u) =
\boldsymbol C : \boldsymbol \epsilon(\boldsymbol u)$ and is composed of the
action of the (possibly anisotropic) linear elasticity tensor $\boldsymbol C$
on the linearised strain $\boldsymbol \epsilon(\boldsymbol u) =
(\nabla \boldsymbol u + \nabla \boldsymbol u^T)/2$.
The solid mass densities inside the volume are denoted by $\rho_s$.
Acting internal volume forces are denoted by $\boldsymbol f$.
A detailed derivation of the system can be found in \cite{koecher_mini_22:Koecher2015}.

Standard notation for function spaces, norms and inner products is used.
Let $\boldsymbol H = L^2(\Omega)^d$, $\boldsymbol V = H^1_0(\Gamma_D;\Omega)^d$,
$\boldsymbol V^* = H^{-1}(\Gamma_D;\Omega)^d$,
$\mathcal H = L^2(I; \boldsymbol H)$,
$\mathcal V = \{ \boldsymbol v \in L^2(I; \boldsymbol V) \,|\,
\partial_t \boldsymbol v \in \mathcal H \}$ and
$\mathcal W = L^2(I; \boldsymbol V)$.
The initial-boundary value problem
with purely homogeneous Dirichlet boundary condition $\boldsymbol g = \boldsymbol 0$
admits an unique weak solution
$\boldsymbol u \in \mathcal V \cap C(\widebar I; \boldsymbol V)$,
$\boldsymbol v \in \mathcal H \cap C(\widebar I; \boldsymbol H)$ and
$\partial_t \boldsymbol v \in L^2(I;\boldsymbol V^*)$ as given by the literature.

A variational space-time discretisation is briefly presented in the sequel;
for all details consider \cite{koecher_mini_22:Koecher2015}.
A piecewise polynomial continuous Galerkin approximation of polynomial degree $r$
is used as semi-discretisation in time. Hereby a piecewise discontinuous space
of polynomial degree $r-1$ is used as test function space which allows the
truncation of the global space-time system to time interval or time slab problems.
Let $0 = t_0 < \dots < t_N = T$ the partition of the temporal domain $I=(0,T)$
into $N$ subintervals $I_n=(t_{n-1}, t_n)$, $n=1,\dots,N$. The length of the
subinterval $I_n$ is defined by $\tau_n=t_n-t_{n-1}$ and let
$\tau = \max_{1\le n \le N} \tau_n$ the global time discretisation parameter.
Let $\mathbb{P}_r(I_n; X)$ denote the space of polynomials of degree $r$ or less
on the interval $I_n \subset I$ with values in some Banach space $X$.
Introducing
$\xi \in C(\widebar I;\mathbb{R})$,
with $\xi|_{I_n} \in \mathbb{P}_r(\widebar I_n;\mathbb{R})$, and
$\zeta \in L^2(I;\mathbb{R})$,
with $\zeta|_{I_n} \in \mathbb{P}_r(\widebar I_n;\mathbb{R})$,
as global piecewise polynomial trail and test basis functions.
The semi-discrete system reads as: Find the pairs
$\{ {\boldsymbol u}_{\tau}^{\textnormal{cG}}|_{I_n},
{\boldsymbol v}_{\tau}^{\textnormal{cG}}|_{I_n} \} \in
\mathbb{P}_r(I_n;\boldsymbol V) \times \mathbb{P}_r(I_n;\boldsymbol V)$,
with coefficients
${\boldsymbol U}_n^{\iota}, {\boldsymbol V}_n^{\iota} \in \boldsymbol V$
for $\iota=1,\dots,r$, such that
\begin{equation}
\label{koecher_mini_22:eq:2:semidiscrete}
\begin{array}{l@{\,}c@{\,}r@{\,}c@{\,}l@{\quad}l}
\displaystyle \sum_{\iota=0}^r \Big\{
\alpha_{\kappa,\iota}\, (\rho_s\, {\boldsymbol U}_n^{\iota}, \widetilde{\boldsymbol \omega})
&-& \beta_{\kappa,\iota}\, (\rho_s\, {\boldsymbol V}_n^{\iota}, \widetilde{\boldsymbol \omega}) \Big\}
&=& \boldsymbol 0\,, &
\forall \widetilde{\boldsymbol \omega} \in \boldsymbol V\,,\\[1.5ex]
\displaystyle \sum_{\iota=0}^r \Big\{
\alpha_{\kappa,\iota}\, (\rho_s\, {\boldsymbol V}_n^{\iota}, \widehat{\boldsymbol \omega})
&+& \beta_{\kappa,\iota}\, a({\boldsymbol U}_n^{\iota}, \widehat{\boldsymbol \omega}) \Big\}
&=&
\displaystyle \sum_{\iota=0}^r
\beta_{\kappa,\iota}\, ({\boldsymbol F}_n^{\iota}, \widehat{\boldsymbol \omega})\,, &
\forall \widehat{\boldsymbol \omega} \in \boldsymbol V\,,
\end{array}
\end{equation}
for all $\kappa=1,\dots,r$, and with
\begin{equation}
\label{koecher_mini_22:eq:3:time_assemblies}
\begin{array}{r@{\,}c@{\,}l@{\,}c@{\,}l@{\,}c@{\,}l}
\alpha_{\kappa,\iota} &:=& (\xi_{n,\iota}^{\prime}(t), \zeta_{n,\kappa}(t))_{I_n} &=&
\displaystyle \int_{\widehat{I}} \widehat \xi_{\iota}^{\prime}(\widehat t)\,
\widehat \zeta_{\kappa}(\widehat t)\, \mathrm{d} \widehat t &=&
\displaystyle \sum_{\mu=0}^{r} \widehat w_{\mu}\,
\widehat \xi_{\iota}^{\prime}(\widehat t_{\mu})\,
\widehat \zeta_{\kappa}(\widehat t_{\mu})\,,\\[1.5ex]
\beta_{\kappa,\iota} &:=& (\xi_{n,\iota}(t), \zeta_{n,\kappa}(t))_{I_n} &=&
\displaystyle \int_{\widehat{I}} \widehat \xi_{\iota}(\widehat t)\,
\widehat \zeta_{\kappa}(\widehat t)\, \tau_n\, \mathrm{d} \widehat t &=&
\displaystyle \sum_{\mu=0}^{r} \tau_n\, \widehat w_{\mu}\,
\widehat \xi_{\iota}(\widehat t_{\mu})\,
\widehat \zeta_{\kappa}(\widehat t_{\mu})\,,
\end{array}
\end{equation}
using quadrature weights $\widehat w_{\mu}$ and quadrature points
$\widehat t_{\mu}$ from $\widehat{\mathcal Q}_{\text{GL($r$+1)}}$
such that the continuity conditions between neighbouring subintervals hold.

For the discretisation in space, let the partition $\mathcal T_h$
of the space domain $\Omega$ into some finite number of disjoint elements $K$.
Denoting by $h_K$ the diameter of the element $K$ and
the global space discretisation parameter as $h = \max_{K \in \mathcal T_h} h_K$.
The spatial mesh is allowed to be anisotropic but non-degenerated.
Let the partition $\widebar{\mathcal F}_h =
\widebar{\mathcal F}_h^{\mathcal B} \cup \mathcal F_h^{\mathcal I}$
into disjoint parts $\widebar{\mathcal F}_h^{\mathcal B}$,
$\mathcal F_h^{\mathcal I}$
for the set of boundary faces and the set of interior faces.
The set of boundary faces is divided as
$\widebar{\mathcal F}_h^{\mathcal B} :=
\mathcal F_h^{\mathcal B(\Gamma_D)} \cup
\mathcal F_h^{\mathcal B(\Gamma_N)}$,
$\mathcal F_h^{\mathcal B} :=
\mathcal F_h^{\mathcal B(\Gamma_D)}$,
into disjoint parts $\mathcal F_h^{\mathcal B(\Gamma_D)}$,
$\mathcal F_h^{\mathcal B(\Gamma_N)}$ coinciding with the Dirichlet
boundary $\Gamma_D$ and the Neumann boundary $\Gamma_N$.
We let $\mathcal F_h :=
\mathcal F_h^{\mathcal B(\Gamma_D)} \cup \mathcal F_h^{\mathcal I} =
\mathcal F_h^{\mathcal B} \cup \mathcal F_h^{\mathcal I}$.
The jump trace operators, which enforce weak Dirichlet boundary conditions,
are defined by
\begin{equation}
\label{koecher_mini_22:eq:4:jump}
\begin{array}{r @{\quad} r}
\jump{\boldsymbol v}_0 :=
\begin{cases}
\boldsymbol v|_{F^+} - \boldsymbol v|_{F^-}\,,
& F \in \mathcal{F}_h^{\mathcal{I}}\,,\\[.5ex]
\boldsymbol v|_{F^+}\,,
& F \in \mathcal{F}_h^{\mathcal{B}}\,,
\end{cases} &
\jump{\boldsymbol v} :=
\begin{cases}
\boldsymbol v|_{F^+} - \boldsymbol v|_{F^-}\,,
& F \in \mathcal{F}_h^{\mathcal{I}}\,,\\[.5ex]
\boldsymbol v|_{F^+} - \boldsymbol v|_{\Gamma_D}\,,
& F \in \mathcal{F}_h^{\mathcal{B}}\,.
\end{cases}
\end{array}
\end{equation}
The average trace operator is defined by
\begin{equation}
\label{koecher_mini_22:eq:5:average}
\average{{\boldsymbol t}_F(\boldsymbol v)} :=
\begin{cases}
\frac{1}{2}\big(
  {\boldsymbol t}_F\big(\boldsymbol v|_{F^+}\big) +
  {\boldsymbol t}_F\big(\boldsymbol v|_{F^-}\big)
  \big)\,, & F \in \mathcal{F}_h^{\mathcal{I}}\,,\\[.5ex]
{\boldsymbol t}_F\big(\boldsymbol v|_{F^+}\big)\,, &
F \in \mathcal{F}_h^{\mathcal{B}}\,,
\end{cases}
\end{equation}
using the traction vector
${\boldsymbol t}_F\big(\boldsymbol v|_{F^\pm}\big) :=
{\boldsymbol \sigma}\big(\boldsymbol v|_{F^\pm} \big) \boldsymbol n^\pm$.

To be concise as possible, we give directly the bilinear form $a_h$ corresponding
to the weak Laplacian as
\begin{equation}
\label{koecher_mini_22:eq:6:a_h}
\begin{array}{r@{\,}c@{\,}l}
a_h({\boldsymbol u}_h^{n,\iota}, \boldsymbol \omega_h) &=&
\displaystyle \sum_{K \in \mathcal T_h} \int_K
{\boldsymbol \sigma}({\boldsymbol u}_h^{n,\iota}) :
{\boldsymbol \epsilon}(\boldsymbol \omega_h)
- \displaystyle \sum_{F \in \mathcal F_h} \int_F
\average{{\boldsymbol t}_F({\boldsymbol u}_h^{n,\iota})} \cdot
\jump{\boldsymbol \omega_h}_0\\[1.5ex]
&~&+ \displaystyle \sum_{F \in \mathcal{F}_h} \int_F
\jump{{\boldsymbol u}_h^{n,\iota}} \cdot \Big(
\gamma_F\, \jump{\boldsymbol \omega_h}_0
- S\,\average{{\boldsymbol t}_F(\boldsymbol \omega_h)}
\Big)\,,
\end{array}
\end{equation}
where $S \in \{1,-1,0\}$ denotes the consistency parameter.
The choice $S=1$ leads to SIPG, $S=-1$ to NIPG and $S=0$ to IIPG.
Note that the non-homogeneous Dirichlet boundary terms can be shifted
efficiently to the right hand side and thus only a stiffness matrix for
homogeneous Dirichlet boundaries has to be assembled.
The value of the interior penalty parameter $\gamma_F$ is determined
by an inverse estimate to balance the terms involving numerical fluxes on the
element boundaries $F \in \mathcal F_h$ in Eq. \eqref{koecher_mini_22:eq:6:a_h}
and to ensure coercivity of the bilinear form $a_h$.
We let $\gamma_F = \gamma_0\, \gamma_{F,{\boldsymbol C}}\, \gamma_{F,K}$,
where $\gamma_{F,K}$ denotes a parameter depending only the polynomial
approximation degree $p$ in space and the shape of the elements $K^\pm$,
$\gamma_{F,{\boldsymbol C}}$ denotes a (scalar) parameter depending the material
parameters in $K^\pm$ and $\gamma_0$ denotes an additional tuning parameter;
cf. for details \cite{koecher_mini_22:Koecher2015,koecher_mini_22:Hoppe2008}.

\section{Fully discrete systems}
\label{koecher_mini_22:sec:3}

In this work only the linear systems of continuous piecewise linear approximations
in time, that are SIPG($p$)--cG($1$) and FEM($p$)--cG($1$), are studied. The
results can be used for higher order time discretisations,
since their efficiency depends strongly on solving additional problems
corresponding to the following systems;
cf. for details \cite{koecher_mini_22:Koecher2015}.

To derive fully discrete systems from Eq. \eqref{koecher_mini_22:eq:2:semidiscrete},
the basis functions in time are specified as Lagrange polynomials,
which are defined via the quadrature points $\widehat t_0 = 0$ and
$\widehat t_1 = 1$ of the two-point Gau\ss{}-Lobatto quadrature rule
$\widehat{\mathcal Q}_{\text{GL($2$)}}$ on the time reference interval
$\widehat I=[0,1]$.
\begin{displaymath}
\begin{array}{l@{\,}c@{\,}l@{\quad}l@{\,}c@{\,}l@{\quad\quad\quad}l@{\,}c@{\,}l}
\widehat \xi_0(\widehat t) &=& 1 - \widehat t\,, &
\widehat \xi_0^\prime(\widehat t) &=& -1\,, \\[1.5ex]
\widehat \xi_1(\widehat t) &=& \widehat t\,, &
\widehat \xi_1^\prime(\widehat t) &=& 1\,, &
\widehat \zeta_1(\widehat t) &=& 1\,, 
\end{array}
\end{displaymath}
and, with that, the coefficients $\alpha_{\kappa,\iota}$ and $\beta_{\kappa,\iota}$
from Eq. \eqref{koecher_mini_22:eq:3:time_assemblies} are evaluated as
\begin{displaymath}
\begin{array}{l@{\,}c@{\,}l@{\quad}l@{\,}c@{\,}l}
\alpha_{1,0} &=& -1\,,              &\alpha_{1,1} &=& 1\,,
\end{array}\quad\quad\quad
\begin{array}{l@{\,}c@{\,}l@{\quad}l@{\,}c@{\,}l}
\beta_{1,0} &=& \frac{\tau_n}{2}\,, &\beta_{1,1}  &=& \frac{\tau_n}{2}\,.
\end{array}
\end{displaymath}
We recast the arising algebraic system by the following linear system with
block structure:
Find the coefficient vectors ${\boldsymbol u}_{I_n}^1,
{\boldsymbol v}_{I_n}^1 \in \mathbb{R}^{N_{\textnormal{DoF}}}$ from
$\boldsymbol L\, \boldsymbol x = \boldsymbol b$ given by
\begin{equation}
\label{koecher_mini_22:eq:7:L}
\left[\!\!\begin{array}{r@{\;}r}
-\frac{\tau_n}{2} \boldsymbol M & \boldsymbol M \\[1.5ex]
\boldsymbol M & \frac{\tau_n}{2} \boldsymbol A
\end{array}\!\!\right]
\left[\!\!\begin{array}{c}
{\boldsymbol v}_{I_n}^1\\[1.5ex]
{\boldsymbol u}_{I_n}^1
\end{array}\!\!\right]
= \left[\!\!\begin{array}{c}
\boldsymbol 0\\[1.5ex]
\frac{\tau_n}{2} ( {\boldsymbol b}_{I_n}^0 + {\boldsymbol b}_{I_n}^1 )
\end{array}\!\!\right]
+ \left[\!\!\begin{array}{r@{\;}r}
\frac{\tau_n}{2} \boldsymbol M & \boldsymbol M \\[1.5ex]
\boldsymbol M & -\frac{\tau_n}{2} \boldsymbol A
\end{array}\!\!\right]
\left[\!\!\begin{array}{c}
{\boldsymbol v}_{I_n}^0\\[1.5ex]
{\boldsymbol u}_{I_n}^0
\end{array}\!\!\right]\,,
\end{equation}
and denoting by $\boldsymbol M$ the mass matrix and by $\boldsymbol A$ the
stiffness matrix.
The assemblies ${\boldsymbol b}_{I_n}^0$ and ${\boldsymbol b}_{I_n}^1$
include the contributions from the forcing terms and inhomogeneous Dirichlet
boundary from Eq. \eqref{koecher_mini_22:eq:6:a_h} using
Eq. \eqref{koecher_mini_22:eq:4:jump}.

The flexible GMRES method with an inexact Krylov-preconditioner can
be used for example to solve the linear system \eqref{koecher_mini_22:eq:7:L}
efficiently.
With some algebraic steps, the block system can be condensed to
$\boldsymbol K\, {\boldsymbol u}_{I_n}^1 = \widetilde{\boldsymbol b}$
and a postprocessing as given by
\begin{equation}
\label{koecher_mini_22:eq:8:K}
\begin{array}{r@{\,}c@{\,}l}
(\boldsymbol M + \frac{\tau_n^2}{4} \boldsymbol A)\, {\boldsymbol u}_{I_n}^1 &=&
\frac{\tau_n^2}{4} ( {\boldsymbol b}_{I_n}^0 + {\boldsymbol b}_{I_n}^1 ) +
( \boldsymbol M - \frac{\tau_n^2}{4} \boldsymbol A ) {\boldsymbol u}_{I_n}^0 +
\tau_n \boldsymbol M {\boldsymbol v}_{I_n}^0\,,\\[1.5ex]
{\boldsymbol v}_{I_n}^1 &=& \frac{2}{\tau_n} (
{\boldsymbol u}_{I_n}^1 - {\boldsymbol u}_{I_n}^0 )
- {\boldsymbol v}_{I_n}^0\,.
\end{array}
\end{equation}
The linear system Eq. \eqref{koecher_mini_22:eq:8:K} is comparable to a
classical discretisation by employing a Crank-Nicolson scheme for the time
discretisation.
Hence, optimised solvers for linear systems with the matrix $\boldsymbol K$
can be re-used.

\section{Numerical Experiments}
\label{koecher_mini_22:sec:4}

In this section we study some numerical properties of the linear systems as
of Eq. \eqref{koecher_mini_22:eq:8:K} and Eq. \eqref{koecher_mini_22:eq:7:L}.
Therefore we approximate the analytic solution
\begin{equation}
\label{koecher_mini_22:eq:9:u_E}
\boldsymbol u^E([x_1,x_2]^T, t) =
\left[\begin{array}{c}
\sin((t+x_1)\cdot 2 \pi)\\[1ex]
\sin((t+x_2)\cdot 2 \pi)\\
\end{array}\right]
\end{equation}
on $\Omega \times I = (0,1)^2 \times (0,1)$ with $\partial \Omega = \Gamma_D$.
The right hand side, initial and boundary values are derived by plugging
$\boldsymbol u^E$ into Eq. \eqref{koecher_mini_22:eq:1:ewave}.
The global mesh size is $h=10^{-1}/(2 \sqrt{2})$ and we use $p=2$ elements in
space.
The material is isotropic with Young's E modulus $E=70$, Poisson's ratio
$\nu=0.34$ and density $\rho_s=2.8$.
The numerical simulations are done with the \textsf{DTM++/ewave} front\-end
solver of the author for the \textsf{deal.II} library;
cf. \cite{koecher_mini_22:Koecher2015}.

\begin{figure}[t]
\centering

\includegraphics[width=.3\linewidth]{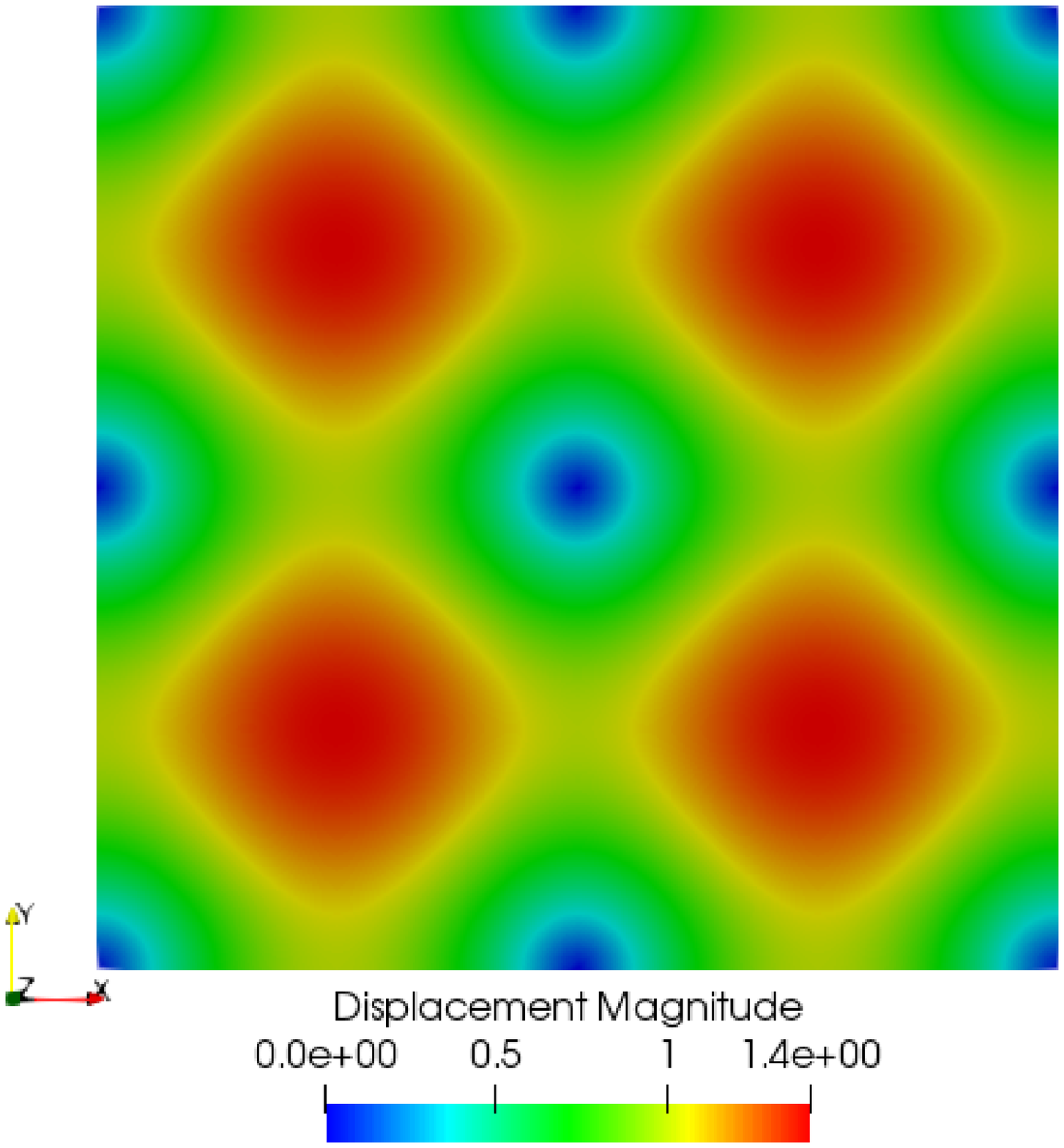}
\includegraphics[width=.3\linewidth]{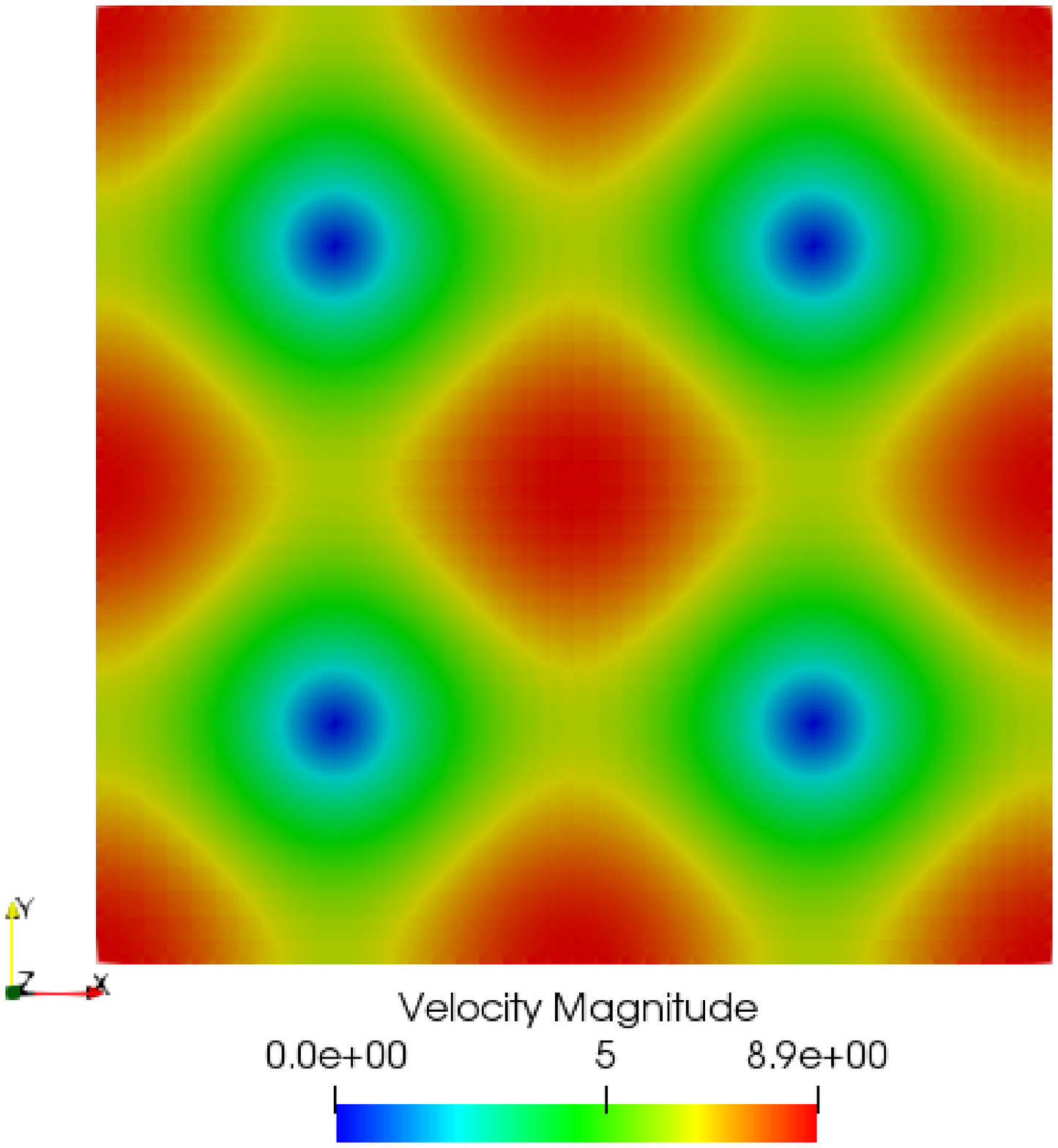}

\caption{Visualisation of the solution Eq. \eqref{koecher_mini_22:eq:9:u_E}
at $t=1$ for the magnitude of displacement (left) and the velocity (right)
with SIPG($2$)--cG($1$) and $\tau_n= 1.25 \cdot 10^{-2}$.}
\label{koecher_mini_22:fig:1:graphical}
\end{figure}

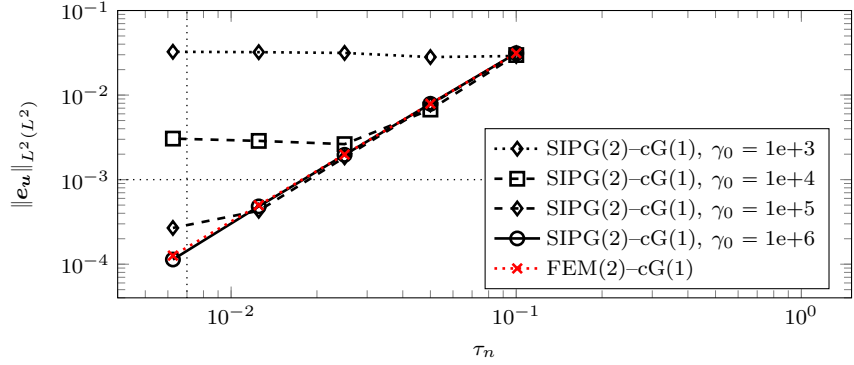
\begin{figure}[p]
\centering
\begin{tikzpicture}
\begin{footnotesize}
\begin{axis}[%
width=3.8in,
height=1.5in,
scale only axis,
xlabel={$\tau_n$},
ylabel={$\| \boldsymbol e_{\boldsymbol u} \|_{L^2(L^2)}$},
xmode=log,
xmin=4e-03,
xmax=1.5e-00,
xminorticks=true,
ymode=log,
ymin=4e-05,
ymax=1e-01,
yminorticks=true,
legend style={legend pos=south east},
legend style={draw=black,fill=white,legend cell align=left}
]

\draw[black,dotted,line width=.5pt]
 (axis cs:1e-03,1e-03) -- (axis cs:1e+00,1e-03);

\draw[black,dotted,line width=.5pt]
 (axis cs:7e-03,1e+00) -- (axis cs:7e-03,1e-08);


\addplot [
color=black,
dotted,
line width=1.0pt,
mark=diamond,
mark size = 2.5,
mark options={solid}
]
table[row sep=crcr]{
1.000e-01 2.8995e-02\\
5.000e-02 2.8154e-02\\
2.500e-02 3.1501e-02\\
1.250e-02 3.2179e-02\\
6.250e-03 3.2506e-02\\
};
\addlegendentry{SIPG(2)--cG(1), $\gamma_0 = \textnormal{1e+3}$};

\addplot [
color=black,
dashed,
line width=1.0pt,
mark=square,
mark size = 2.5,
mark options={solid}
]
table[row sep=crcr]{
1.000e-01 2.9756e-02\\
5.000e-02 6.7419e-03\\
2.500e-02 2.6300e-03\\
1.250e-02 2.8720e-03\\
6.250e-03 3.0568e-03\\
};
\addlegendentry{SIPG(2)--cG(1), $\gamma_0 = \textnormal{1e+4}$};

\addplot [
color=black,
dashed,
line width=1.0pt,
mark=diamond,
mark size = 2.5,
mark options={solid}
]
table[row sep=crcr]{
1.000e-01 3.1136e-02\\
5.000e-02 7.7434e-03\\
2.500e-02 1.8346e-03\\
1.250e-02 4.2826e-04\\
6.250e-03 2.6840e-04\\
};
\addlegendentry{SIPG(2)--cG(1), $\gamma_0 = \textnormal{1e+5}$};

\addplot [
color=black,
solid,
line width=1.0pt,
mark=o,
mark size = 2.5,
mark options={solid}
]
table[row sep=crcr]{
1.000e-01 3.1281e-02\\
5.000e-02 7.8875e-03\\
2.500e-02 1.9658e-03\\
1.250e-02 4.8245e-04\\
6.250e-03 1.1355e-04\\
};
\addlegendentry{SIPG(2)--cG(1), $\gamma_0 = \textnormal{1e+6}$};

\addplot [
color=red,
dotted,
line width=1.0pt,
mark=x,
mark size = 2.5,
mark options={solid}
]
table[row sep=crcr]{
1.000e-01 3.1298e-02\\
5.000e-02 7.9037e-03\\
2.500e-02 1.9822e-03\\
1.250e-02 4.9765e-04\\
6.250e-03 1.2630e-04\\
};
\addlegendentry{FEM(2)--cG(1)};
\end{axis}
\end{footnotesize}
\end{tikzpicture}%

\caption{Interior penalty parameter $\gamma_F$ influence on the
experimental convergence behaviour in time
for approximating $\boldsymbol u^E$ from Eq. \eqref{koecher_mini_22:eq:9:u_E}
for Sec. \ref{koecher_mini_22:sec:4}.}
\label{koecher_mini_22:fig:2:Penalty2Error}
\end{figure}

\begin{figure}[p]
\centering
\begin{tikzpicture}
\begin{footnotesize}
\begin{axis}[%
width=3.8in,
height=1.5in,
scale only axis,
xlabel={$\tau_n$},
ylabel={experimental condition number},
xmode=log,
xmin=1.7e-06,
xmax=1.8e-01,
xminorticks=true,
ymode=log,
ymin=1e+00,
ymax=1e+08,
yminorticks=true,
legend style={legend pos=north west},
legend style={draw=black,fill=white,legend cell align=left}
]

\draw[black,densely dotted,line width=.5pt]
 (axis cs:7e-03,1e+00) -- (axis cs:7e-03,1e+08);

\draw[fill=black!5!white,draw=none]
 (axis cs:7.1e-03,2e+00) rectangle (axis cs:1.4e-01,4e+07);

\node at (axis cs:3.2e-02,1e+01) {low accuracy};

\addplot [
color=black,
solid,
line width=1.0pt,
mark=o,
mark size = 2.5,
mark options={solid}
]
table[row sep=crcr]{
 1.0000e-01 5.547993605064002e+06\\
 5.0000e-02 4.689081024486233e+06\\
 2.5000e-02 2.910044080073575e+06\\
 1.2500e-02 1.193517330389460e+06\\
 6.2500e-03 0.418749301745966e+06\\
 3.1250e-03 0.202948604879645e+06\\
 1.5625e-03 0.120627461782131e+06\\
 7.8125e-04 0.058641631968293e+06\\
 3.9063e-04 0.020268961073798e+06\\
 1.9531e-04 0.005641446498536e+06\\
 9.7656e-05 0.001459069296600e+06\\
 4.8828e-05 0.000375300082320e+06\\
 2.4414e-05 0.000103412473667e+06\\
 1.2207e-05 0.000050683185368e+06\\
 6.1035e-06 0.000047785452364e+06\\
 3.0518e-06 0.000047424092873e+06\\
};
\addlegendentry{SIPG(2)--cG(1), $\gamma_0 = \textnormal{1e+6}$};

\addplot [
color=black,
dashed,
line width=1.0pt,
mark=diamond,
mark size = 2.5,
mark options={solid}
]
table[row sep=crcr]{
 1.0000e-01 1.156411857174727e+07\\
 5.0000e-02 0.238391485365007e+07\\
 2.5000e-02 0.057647134492397e+07\\
 1.2500e-02 0.014873233712127e+07\\
 6.2500e-03 0.004466257915990e+07\\
 3.1250e-03 0.002088178786071e+07\\
 1.5625e-03 0.001228029055532e+07\\
 7.8125e-04 0.000592133986911e+07\\
 3.9063e-04 0.000204104475165e+07\\
 1.9531e-04 0.000057356265419e+07\\
 9.7656e-05 0.000015591278385e+07\\
 4.8828e-05 0.000005613608928e+07\\
 2.4414e-05 0.000004813307108e+07\\
 1.2207e-05 0.000004748168384e+07\\
 6.1035e-06 0.000004736438914e+07\\
 3.0518e-06 0.000004733829807e+07\\
};
\addlegendentry{SIPG(2)--cG(1), $\gamma_0 = \textnormal{1e+5}$};



\addplot [
color=red,
dotted,
line width=1.0pt,
mark=x,
mark size = 2.5,
mark options={solid}
]
table[row sep=crcr]{
 1.0000e-01 4.101771059113861e+03\\
 5.0000e-02 1.033570762668442e+03\\
 2.5000e-02 0.275524017001339e+03\\
 1.2500e-02 0.098171527856643e+03\\
 6.2500e-03 0.049170081552520e+03\\
 3.1250e-03 0.027720006755321e+03\\
 1.5625e-03 0.019867070379011e+03\\
 7.8125e-04 0.023315907971360e+03\\
 3.9063e-04 0.027113174690742e+03\\
 1.9531e-04 0.028628938252499e+03\\
 9.7656e-05 0.029065174742971e+03\\
 4.8828e-05 0.029178391045071e+03\\
 2.4414e-05 0.029206965498963e+03\\
 1.2207e-05 0.029214126183119e+03\\
 6.1035e-06 0.029215917423792e+03\\
 3.0518e-06 0.029216365300856e+03\\
};
\addlegendentry{FEM(2)--cG(1)};
\end{axis}
\end{footnotesize}
\end{tikzpicture}%

\caption{Influence of the interior penalty parameter $\gamma_F$
and the global time discretisation parameter $\tau_n$
on the experimental condition number for Sec. \ref{koecher_mini_22:sec:4}.}
\label{koecher_mini_22:fig:3:ECNvalue}
\end{figure}
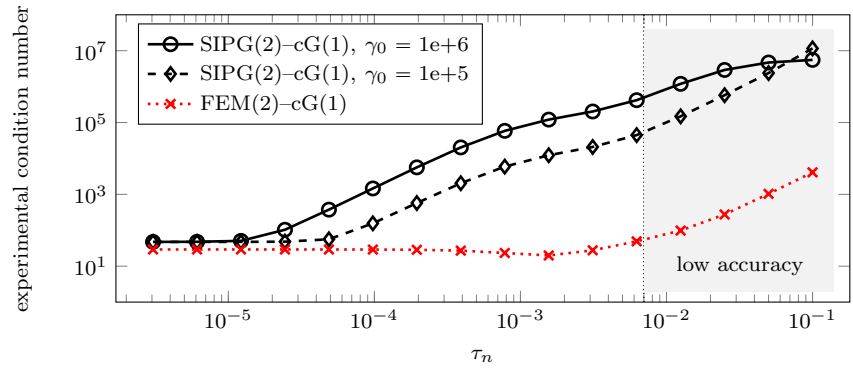

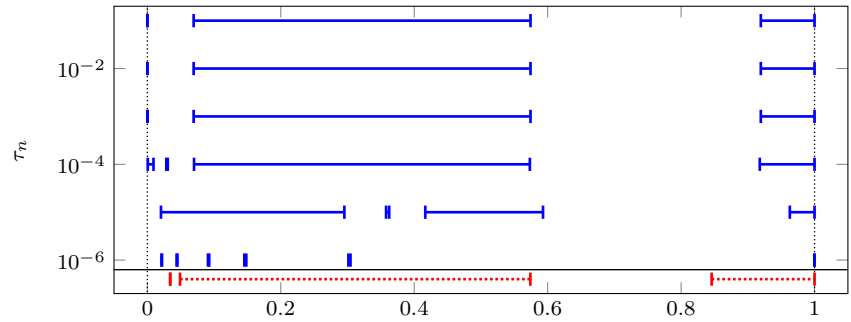
\begin{figure}[p]
\centering
\begin{tikzpicture}
\begin{footnotesize}
\begin{axis}[%
width=3.8in,
height=1.5in,
scale only axis,
ylabel={$\tau_n$},
xmin=-0.05,
xmax=1.05,
xminorticks=true,
ymode=log,
ymin=2e-07,
ymax=2e-01,
yminorticks=false,
legend style={legend pos=south east},
legend style={draw=black,fill=white,legend cell align=left}
]

\draw[black,densely dotted,line width=.5pt]
 (axis cs:0,1e-07) -- (axis cs:0,1);

\draw[black,densely dotted,line width=.5pt]
 (axis cs:1,1e-07) -- (axis cs:1,1);

\addplot [
color=blue,
solid,
line width=1.0pt,
mark=|,
mark size = 2.5,
mark options={solid}
]
table[row sep=crcr]{
0.000000180245341 1e-01\\
0.000073513722524 1e-01\\
};

\addplot [
color=blue,
solid,
line width=1.0pt,
mark=|,
mark size = 2.5,
mark options={solid}
]
table[row sep=crcr]{
0.069354258985711 1e-01\\
0.574291667212066 1e-01\\
};

\addplot [
color=blue,
solid,
line width=1.0pt,
mark=|,
mark size = 2.5,
mark options={solid}
]
table[row sep=crcr]{
0.919502559209207 1e-01\\
1.00 1e-01\\
};

\addplot [
color=blue,
solid,
line width=1.0pt,
mark=|,
mark size = 2.5,
mark options={solid}
]
table[row sep=crcr]{
0.000001179702951 1e-02\\
0.000075317761632 1e-02\\
};

\addplot [
color=blue,
solid,
line width=1.0pt,
mark=|,
mark size = 2.5,
mark options={solid}
]
table[row sep=crcr]{
0.069354320986574 1e-02\\
0.574291572753600 1e-02\\
};

\addplot [
color=blue,
solid,
line width=1.0pt,
mark=|,
mark size = 2.5,
mark options={solid}
]
table[row sep=crcr]{
0.919502408654538 1e-02\\
1.00 1e-02\\
};

\addplot [
color=blue,
solid,
line width=1.0pt,
mark=|,
mark size = 2.5,
mark options={solid}
]
table[row sep=crcr]{
0.000012695294295 1e-03\\
0.000351544463657 1e-03\\
};

\addplot [
color=blue,
solid,
line width=1.0pt,
mark=|,
mark size = 2.5,
mark options={solid}
]
table[row sep=crcr]{
0.069360520740998 1e-03\\
0.574282124676611 1e-03\\
};

\addplot [
color=blue,
solid,
line width=1.0pt,
mark=|,
mark size = 2.5,
mark options={solid}
]
table[row sep=crcr]{
0.919487349619777 1e-03\\
1.00 1e-03\\
};

\addplot [
color=blue,
solid,
line width=1.0pt,
mark=|,
mark size = 2.5,
mark options={solid}
]
table[row sep=crcr]{
0.000654137855591 1e-04\\
0.009218553133667 1e-04\\
};

\addplot [
color=blue,
solid,
line width=1.0pt,
mark=|,
mark size = 2.5,
mark options={solid}
]
table[row sep=crcr]{
0.028409492704049 1e-04\\
0.030583369811418 1e-04\\
};

\addplot [
color=blue,
solid,
line width=1.0pt,
mark=|,
mark size = 2.5,
mark options={solid}
]
table[row sep=crcr]{
0.069977291479839 1e-04\\
0.573314588400109 1e-04\\
};

\addplot [
color=blue,
solid,
line width=1.0pt,
mark=|,
mark size = 2.5,
mark options={solid}
]
table[row sep=crcr]{
0.917945056461247 1e-04\\
1.00 1e-04\\
};

\addplot [
color=blue,
solid,
line width=1.0pt,
mark=|,
mark size = 2.5,
mark options={solid}
]
table[row sep=crcr]{
0.020387665100952 1e-05\\
0.295301364236930 1e-05\\
};

\addplot [
color=blue,
solid,
line width=1.0pt,
mark=|,
mark size = 2.5,
mark options={solid}
]
table[row sep=crcr]{
0.357815096707493 1e-05\\
0.362468687981915 1e-05\\
};

\addplot [
color=blue,
solid,
line width=1.0pt,
mark=|,
mark size = 2.5,
mark options={solid}
]
table[row sep=crcr]{
0.416570911509303 1e-05\\
0.593035083658206 1e-05\\
};

\addplot [
color=blue,
solid,
line width=1.0pt,
mark=|,
mark size = 2.5,
mark options={solid}
]
table[row sep=crcr]{
0.962922202309228 1e-05\\
1.00 1e-05\\
};

\addplot [
color=blue,
solid,
line width=1.0pt,
mark=|,
mark size = 2.5,
mark options={solid}
]
table[row sep=crcr]{
0.021124098149327 1e-06\\
0.021985058153612 1e-06\\
};

\addplot [
color=blue,
solid,
line width=1.0pt,
mark=|,
mark size = 2.5,
mark options={solid}
]
table[row sep=crcr]{
0.043802366299623 1e-06\\
0.045144221479497 1e-06\\
};

\addplot [
color=blue,
solid,
line width=1.0pt,
mark=|,
mark size = 2.5,
mark options={solid}
]
table[row sep=crcr]{
0.090827051614494 1e-06\\
0.092689951886727 1e-06\\
};

\addplot [
color=blue,
solid,
line width=1.0pt,
mark=|,
mark size = 2.5,
mark options={solid}
]
table[row sep=crcr]{
0.145322382179391 1e-06\\
0.148303031820215 1e-06\\
};

\addplot [
color=blue,
solid,
line width=1.0pt,
mark=|,
mark size = 2.5,
mark options={solid}
]
table[row sep=crcr]{
0.301335670257221 1e-06\\
0.304464816757817 1e-06\\
};

\addplot [
color=blue,
solid,
line width=1.0pt,
mark=|,
mark size = 2.5,
mark options={solid}
]
table[row sep=crcr]{
0.999742452706679 1e-06\\
1.00 1e-06\\
};

\draw[black,solid,line width=.5pt]
 (axis cs:-1,6.3e-07) -- (axis cs:2,6.3e-07);

\addplot [
color=red,
densely dotted,
line width=1.0pt,
mark=|,
mark size = 2.5,
mark options={solid}
]
table[row sep=crcr]{
0.034227236287898 4e-07\\
0.034227248385157 4e-07\\
};

\addplot [
color=red,
densely dotted,
line width=1.0pt,
mark=|,
mark size = 2.5,
mark options={solid}
]
table[row sep=crcr]{
0.048954602112973 4e-07\\
0.574078950989747 4e-07\\
};

\addplot [
color=red,
densely dotted,
line width=1.0pt,
mark=|,
mark size = 2.5,
mark options={solid}
]
table[row sep=crcr]{
0.845751305574309 4e-07\\
1.00 4e-07\\
};

\end{axis}
\end{footnotesize}
\end{tikzpicture}%

\caption{Distribution of normalised eigenvalues of matrices $\boldsymbol K(\tau_n)$
for SIPG($2$) with $\gamma_0=10^6$ (solid blue) and
for FEM($2$) with $\tau_n=10^{-6}$ (dotted red)
for Sec. \ref{koecher_mini_22:sec:4}.}
\label{koecher_mini_22:fig:4:spectrum}
\end{figure}

The solutions of the displacement and velocity are illustrated
by Fig. \ref{koecher_mini_22:fig:1:graphical}.
The convergence for the error $\boldsymbol e_{\boldsymbol u} =
\boldsymbol u^E - {\boldsymbol u}_{\tau,h}^{\textnormal{cG(1)}}$ and the
dependency on the penalisation $\gamma_0$ to obtain a certain accuracy is
presented in Fig. \ref{koecher_mini_22:fig:2:Penalty2Error}.
For a penalty value of $\gamma_0 = 10^{6}$ the experimental order of convergence
in time is $2.00$ in the $L^2(I;L^2(\Omega))$-norm and
the calculated errors of the SIPG and FEM discretisations are comparable.
Fig. \ref{koecher_mini_22:fig:3:ECNvalue} presents the effects of the penalisation
and the time step length $\tau_n$ on the condition numbers.
Fig. \ref{koecher_mini_22:fig:4:spectrum} illustrates the distribution of the
normalised eigenvalues of the matrix $\boldsymbol K$ for SIPG($2$)--cG($1$) with
$\gamma_0=10^6$ for several values of $\tau_n$ and
the normalised eigenvalues of the matrix $\boldsymbol K$ for FEM($2$)--cG($1$)
for $\tau_n=10^{-6}$.

\section{Conclusions}
\label{koecher_mini_22:sec:5}

The influence of the penalisation of the SIPG discretisation is analysed
numerically for fully discrete linear systems of different representations
and compared with their standard finite element counterparts.
It is shown that the condition number of the condensed SIPG system matrix scales
with the penalisation and the time subinterval length $\tau_n$ with a challenging
numerical experiment.
The eigenvalues collects in small number of clusters for SIPG discretisations
by choosing $\tau_n \approx 1/\gamma_0$.
The effect of clustering of the eigenvalues could not be reproduced for the
comparable FEM system or for the block system with SIPG.
A (much) faster convergence behaviour of the conjugate gradient method for
the condensed SIPG system for small time step sizes was noticed but not analysed
by the author in the past for several three dimensional problems with physically
relevance; cf. \cite{koecher_mini_22:Koecher2015}.
With the results of this work, a faster convergence behaviour of the conjugate
gradient method can be explained due to the clustering effect of the eigenvalues
in such cases.
The results of this work clearly help to design preconditioners for
the block system or the condensed system for SIPG discretisations for the
elastic wave equation, but we keep this as future work.

\section*{Acknowledgements}

The research contribution of the author was partially supported by
E.ON Stipendienfonds (Germany) under the grant T0087 / 29890 / 17 while visiting
University of Bergen.

\def\cprime{$'$}

\ifx\undefined\bysame
\newcommand{\bysame}{\leavevmode\hbox to3em{\hrulefill}\,}
\fi

\end{document}